\documentclass[aps,prl,reprint,twocolumn,floatfix,
superscriptaddress,
citecolor=blue,
linkcolor=blue,
urlcolor=blue,
colorlinks]{revtex4-2}

\usepackage{graphicx}
\usepackage{dcolumn}
\usepackage[normalem]{ulem}
\usepackage{xcolor}
\usepackage{comment}
\usepackage{amsmath, amssymb}
\usepackage{siunitx}
\usepackage{dsfont}
\usepackage{placeins}
\usepackage[shortlabels]{enumitem}
\usepackage{soul}
\usepackage{booktabs}
\usepackage{hyperref}
\usepackage{bm}

\DeclareSIUnit\angstrom{\text{Å}}

\NewDocumentCommand{\figref}{m o}{%
    Fig.~\ref{#1}%
    \IfValueT{#2}{(#2)}%
}

\newcommand{\refcite}[1]{Ref.~\cite{#1}}

\usepackage[columnwise]{lineno}

\begin{document}

\title{Large pseudo-orbital magnetic moment of magnons in $\mathrm{CrI}_3$ nanotubes}

\author{David Sanz}
\affiliation{Departamento de F\'isica, Universidad de Alicante, 03690 San Vicente del Raspeig, Spain}

\author{J. Fern\'andez-Rossier}
\affiliation{Departamento de F\'isica, Universidad de Alicante, 03690 San Vicente del Raspeig, Spain}
\affiliation{International Iberian Nanotechnology Laboratory (INL), Av. Mestre Jos\'e Veiga, 4715-330 Braga, Portugal}

\begin{abstract}
Recently discovered $\mathrm{CrI}_3$ nanotubes are expected to host unusual radial magnetic order states, on account of the strong out-of-plane magnetic anisotropy of the two-dimensional parent material. The  detection of these  peculiar magnetic states is challenging, and motivates the quest for unique features  that can be probed experimentally. Using linear spin-wave theory, we show that an axial magnetic field tilts the radial phase into a conical one that threads the tube with an Aharonov-Bohm flux for magnons. This flux endows magnons with a pseudo-orbital magnetic moment, reaching approximately six Bohr magnetons, three times  conventional magnon spin moment, and therefore detectable through magnon spectroscopy, providing a smoking gun for the radial and conical magnetic order in $\mathrm{CrI}_3$ nanotubes. 
\end{abstract}

\maketitle

\textit{Introduction and motivation}---
$\mathrm{CrI}_3$ is a key compound  in the field of van der Waals magnets, being the first system that allowed for the  experimental observation\cite{huang2017} of ferromagnetic order down to a single monolayer. This triggered a new research branch in the field of van der Waals materials, the so called 2D magnets\cite{Gong2017,Burch2018,Gibertini2019}.
One of the outstanding features of $\mathrm{CrI}_3$ is the strong out-of-plane magnetic anisotropy\cite{lado2017ani,soriano2020review}, which breaks spin rotational invariance, opens a gap in the magnon spectrum and provides stability to the long-range order, evading  of applicability of the Mermin-Wagner theorem\cite{mermin1966}.

\begin{figure}[ht]  
    \centering
    \includegraphics[width=1\linewidth]{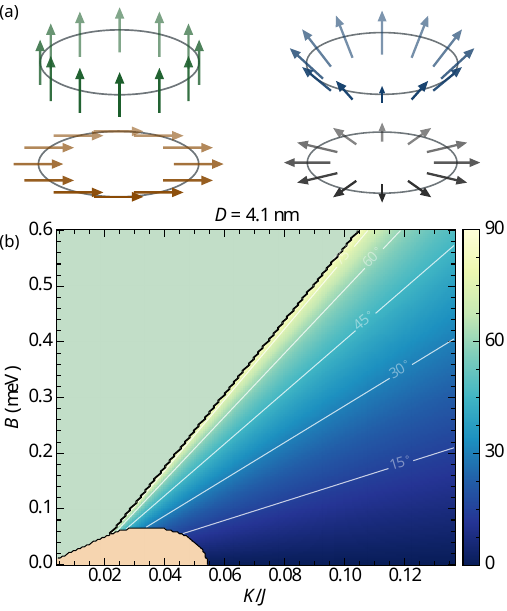}
    \caption{(a) Artistic representation of the four magnetic orders considered: axial ferromagnetic (axial-FM, upper left), conical (upper right), radial (lower right), and transverse ferromagnetic (transverse-FM, lower left). (b) Classical ground-state phase diagram of an $(11,11)$ nanotube with diameter of $4.1$~nm, as a function of the single-ion anisotropy $K$ and axial external magnetic field $B$, obtained by minimizing the classical energy of Eq.~\eqref{eq:spin-ham}: axial-FM (green), transverse-FM (orange), and conical (blue). The latter containing the radial magnetic order as the special case $g \mu_B B=0$~meV. The color gradient gives the conical angle (see colorbar). Contour lines (white solid) give paths in the parameter space with constant conical angle. Fixed parameters are $J = 2.2$~meV, $S=\tfrac{3}{2}$, and lattice parameter of $a = 6.7$~\AA.}
    \label{fig:1}
\end{figure}

The electronic and structural properties of both carbon\cite{dresselhaus98} and inorganic\cite{tenne06} nanotubes have been explored for several decades now. The recent discovery of single-wall nanotubes made of ferromagnetic chromium trihalides, 
$\mathrm{CrI}_3$~\cite{caha2025} and $\mathrm{CrCl}_3$~\cite{ul26}, encapsulated inside multiwall carbon nanotubes, has opened a new branch of research: van der Waals magnetic nanotubes. Whereas so far the magnetic properties of individual $\mathrm{CrI}_3$ tubes have not been measured, both DFT calculations \cite{edstrom2022curved} and theoretical models \cite{caha2025}  predict that, above a critical value for the tube radius $R$, the magnetic ground state is radial. This is to be expected as, in the $R\rightarrow\infty$ limit the out-of-plane easy axis has to be obtained. In addition, in this work we show that application of a magnetic field along the direction of the tube results in a tilted radial phase (see Fig.~\ref{fig:1}) that we dub conical phase.

The study of magnetism, and the resulting magnons, in curved nanostructures has focused mostly on metallic nanostructures, where the radius of curvature is much larger than in the system studied here\cite{balhorn10,gaididei14,sheka2015,umetsu2025}. Recent work has paid attention to the fate of magnons in magnetic nanotubes \cite{jin26, salazar2021nonreciprocity, otalora2017asymmetric, jia2019, nunes2026curvature}. In addition, even in non-curved structures,  non-coplanar and non-collinear magnetic states, such as helices, vortices\cite{dugaev05, schultheiss2019, toth2015, huang2022topological} and skyrmions host magnons with non-trivial properties, such as anomalous velocities, topological gaps and enhanced stability due to quantum fluctuations\cite{roldan2015}.

Probing magnetism down to the monolayer limit has been a key challenge in the field of  two-dimensional van der Waals magnets, on account of the fact that the  SQUID and neutron scattering sensitivity would require unrealistically large monolayers\cite{mak2019probing}. This problem is even worse in the case of van der Waals nanotubes. Therefore, progress in this emerging field will require new ideas on how to probe the exotic magnetic states predicted by theory. With this background, here we show that the interplay between  curvature and out-of-plane magnetic anisotropy induces non-collinear magnetic ground states that endow the resulting magnons with a large pseudo-orbital magnetic moment, that can be detected in local spin resonance\cite{thiel2019probing} and inelastic electron tunneling spectroscopy\cite{klein18,Ghazaryan2018}.

\textit{Model and classical ground states}---
We consider a model Hamiltonian including Heisenberg exchange, single-ion anisotropy  (SIA) and
a Zeeman coupling,
\begin{equation}\label{eq:spin-ham}
H =
-\frac12\sum_{ij} J_{ij}\,\mathbf{S}_i\cdot\mathbf{S}_j
-
K\sum_i
(\mathbf{S}_i\cdot\mathbf{e}_{r,i})^2
-g\mu_B
\sum_i \mathbf{B}\cdot\mathbf{S}_i ,
\end{equation}
where $J_{ij}$, $K$ and $g \mu_B \bm{B}$ are given in energy units. We take $g \simeq 2$, so $g\mu_B \simeq 0.116$~meV/T. The spins reside on the sites of a nanotube aligned along the $x$ axis, defined with the standard notation $(n,m)$, with $n,m$ integer numbers \cite{okuyama2019topological}.

The first term represents the isotropic exchange coupling between nearest-neighbor sites $i$ and $j$, which for $J_{ij}>0$ is ferromagnetic and favors parallel alignment of neighboring spins. The second term is the SIA of strength $K>0$, whose local easy axis is set by the unit vectors
\begin{equation}\label{eq:sia-vector}
\bm{e}_{r,i} = \left(0,\,\cos\varphi_i,\,\sin\varphi_i\right) ,
\end{equation}
with $\varphi_i$ the azimuthal angle of site $i$ around the tube axis, so that $\bm{e}_{r,i}$ points along the radial direction of the tube favoring a radial magnetic order. The third term is the Zeeman coupling to a uniform field $g \mu_B B$ applied along the tube in units of meV, which promotes alignment of the spins along it.

Hence, the model is completely defined once the values of the nearest-neighbor exchange $J$, single-ion anisotropy $K$, magnitude of the longitudinal field $B$, and tube geometry $(n,m)$ are given.  For $\mathrm{CrI}_3$, reasonable parameters are $J=2.2$~meV, $K = 0.13$~meV ($K/J \simeq0.06$) \cite{lado2017ani}. We consider $\mathrm{CrI}_3$ armchair nanotubes with $n=m=9$--$15$, which for a lattice constant of $a = 6.7$~\AA{} corresponds to tube diameters $D \simeq 3.3$--$5.5$~nm, within the values reported in recent experimental realizations \cite{caha2025}.

We describe magnons using conventional Holstein-Primakoff (HP) transformation within linear spin-wave theory (LSWT) \cite{holstein1940}. In this approach, the starting point is the determination of a broken symmetry classical ground state, treating the spin operators in Eq.~\eqref{eq:spin-ham} classically.  We consider four classes of magnetic  ground state: 
radial state, in which each spin points along the radial direction within the transverse plane to the tube; conical state, in which each spin points along its local radial direction and is tilted by a conical angle $\theta$, measured from the transverse plane toward the tube; axial ferromagnetic (axial-FM) state, with all spins pointing parallel to the tube axis; and a transverse ferromagnetic (transverse-FM) state, with each spin parallel to a reference direction within the transverse plane.

The first three phases can be described by the following ansatz for the magnetization unit vector at site $i$:
\begin{equation}
\label{eq:param-magn}
    \bm{m}_i =\cos(\theta) \bm{e}_{r,i} + \sin(\theta) \bm{e}_z,
\end{equation}
with $\bm{e}_{r,i}$ the radial unit vector and $\bm{e}_z$ is the axial unit vector along the tube axis. As shown in Fig.~\ref{fig:1}(a), $\theta=0$ ($\theta=\frac{\pi}{2}$) in Eq.~\eqref{eq:param-magn} corresponds to the radial (axial-FM) solution. As the field $B$ drives the conical angle $\theta$ from 0 to $\frac{\pi}{2}$, the system passes through the conical phase. Minimizing the classical energy of Eq.~\eqref{eq:spin-ham} with the ansatz of Eq.~\eqref{eq:param-magn} as a function of $\theta$ gives
\begin{equation}
\label{eq:bs}
\sin\theta = \frac{B}{B_s}, \quad  g\mu_B B_s\equiv S(2K - J\Gamma_n) ,
\end{equation}
valid for $0 \leq B \leq B_s$, where $B_s$ is the  field above which the system saturates in the axial-FM phase. $\Gamma_n = 3 - \cos\left(\frac{2\pi}{3n}\right) - 2\cos\left(\frac{\pi}{3n}\right)$ for a $(n,n)$ armchair tube, which yields in the wide-tube limit $g \mu_B B_s \to 2KS$.

We assume that magnetic solutions respect the translational invariance of the tube, neglecting the formation of domain walls \cite{edstrom2022curved}. The validity of these solutions as local minima is verified by the fact that the energies of the magnons are real and positive~\cite{colpa1978, toth2015}. The exploration of the existence of other classical solutions with lowest energy is out of the scope of this work.

The classical ground state for each case is determined by comparing the energies of the transverse-FM phase and the ansatz given in Eq.~\eqref{eq:param-magn}.\ Fig.~\ref{fig:1}(b) shows the phase diagram in the ($K/J, B$) plane for the parameters given in the caption.  At $B=0$, only two phases appear: transverse-FM and radial ($\theta=0$). The critical value of $K/J$ that separates them depends on the tube diameter~\cite{caha2025}, as the exchange energy overhead of the radial solutions is larger for smaller diameter. A moderate field $B$ applied along the tube axis tilts the spins of the radial solution toward the axis, yielding a conical ground state. When $B$ is large enough, the spins are fully aligned along the tube axis. By contrast, at small $K/J$ and $B$, the transverse-FM phase is the ground state (orange region in Fig.~\ref{fig:1}). 
As the tube diameter increases, the transverse-FM region shrinks in the phase diagram and the radial phase becomes dominant at zero field and wide range of $K/J$.  For the parameters of $\mathrm{CrI}_3$, we use $K = 0.13$~meV ($K/J \simeq 0.06$), so the ground state of tubes with $n\ge 11$ lies outside the transverse-FM region. We therefore restrict our analysis to the radial, conical and axial-FM phases.

\textit{Magnon Hamiltonian---}
The derivation of the magnon Hamiltonian within LSWT follows the standard procedure for noncollinear magnets \cite{colpa1978, toth2015,roldan2015} (see  End Matter A). The spin operators in Eq.~\eqref{eq:spin-ham} are expressed in terms of Holstein-Primakoff (HP) bosons~\cite{holstein1940}, taking as the local quantization axis the direction of the classical ground state. The Hamiltonian is then truncated at quadratic order in the HP bosonic operators. Translational invariance along the axial direction is exploited to write the Hamiltonian as a sum over wave vectors $k$ of $2N_c \times 2N_c$ Bogoliubov de Gennes (BdG) matrices $h(k)$, where $N_c$ is the number of spins in the unit cell. Using the translational symmetry of the $N_c = 4n$ axial unit cell, we can Fourier transform the bosonic operators and label them by the axial wave vectors $k$. We refer to this approach as the axial representation.

In addition, we take advantage of the rotational symmetry $C_n$ and the helical symmetry of the $(n,n)$ tubes, the latter generated by a screw operation that combines a rotation by $\pi/n$ with a translation by $a/2$ along the tube axis~\cite{okuyama2019topological}. By doing so, we can label the Hamiltonian sectors by the angular momentum $\mu=0,\dots,n-1$ and the helical wave vector $q$, conjugate to the screw operation. In this approach, given a $(\mu, q)$ pair, the magnon dispersion is obtained by diagonalizing the corresponding $4\times4$ BdG matrix. We refer to this approach as the helical representation.

In both the axial and the helical representations, we introduce a paraunitary transformation~\cite{colpa1978} to diagonalize the BdG matrices, which maps the HP bosons into BdG quasiparticles, $\alpha_{m,\bm{k}}$, while preserving the bosonic commutation relations.  In the following, we refer to the BdG quasiparticles as magnons. Applying this process to Eq.~\eqref{eq:magnon-ham-quadratic} and up to a c-number term, the Hamiltonian in diagonal form is given by,
\begin{equation}
\label{eq:magnon-ham-final}
    H = \sum_{m,\bm{k}} E_m(\bm{k}) \alpha^\dagger_{m,\bm{k}} \alpha_{m,\bm{k}}
\end{equation}
where $E_m(\bm{k})$ is the magnon dispersion energy for the $m$ magnon band.

\begin{figure}[t]
    \centering
    \includegraphics[width=\linewidth]{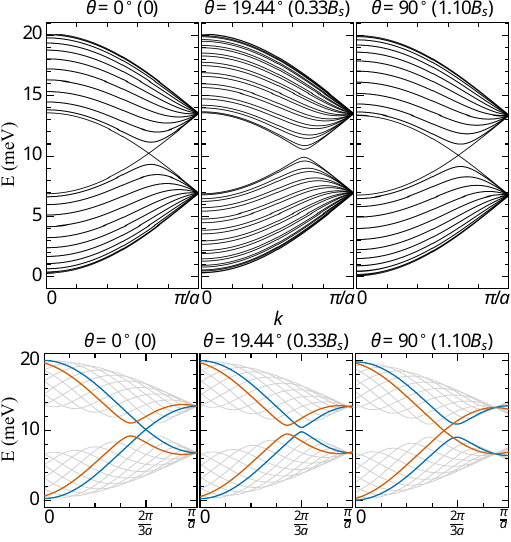}
    \caption{Magnon dispersion of the $(11-11)$ armchair nanotube for $J=2.2$~meV, S=$3/2$ and K=$0.13$~meV, in the three phases: radial ($\theta=0^\circ, B=0$), conical ($\theta=19.44^\circ, B=0.33B_s$) and axial-FM ($\theta=90^\circ, B=1.1B_s$). 
    (a) Top-row: bands obtained with the axial unit cell. (b) Bottom-row: the same spectra, but obtained through the helical representation. Each band is labeled by its angular momentum integer $\mu=0,\dots,n-1$ for a $(n,n)$ tube. Blue (orange) solid lines highlight bands with $\mu_D=0$ ($\mu_{D'}=n-1$). All other bands are shown in faded gray. The crossings are formed by the $\mu_D$ bands in the radial phase and $\mu_{D'}$ in the axial-FM phase, and it is gapped in the conical phase. }
    \label{fig:2}
\end{figure}

\begin{figure}[t]
    \centering
    \includegraphics[width=0.9\linewidth]{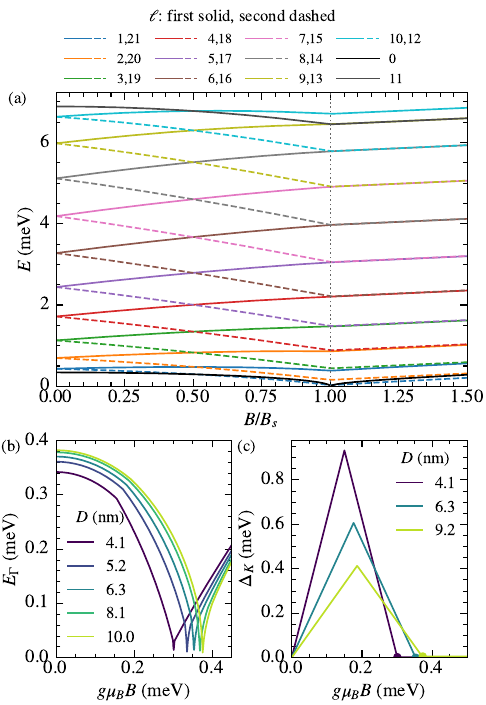}
    \caption{Splittings of magnon energies, $\Delta_{\Gamma}$, and $\Delta_{K}$ of the $(11,11)$ nanotube. (a) Magnon energies at $\Gamma$
    as a function of $B/B_s$, where the slopes give the magnetic moments $m_z=- d E/ d B$.
    Colors label by the angular index $\ell$ the degenerate pairs $(\ell, \bar{\ell})$ and non-degenerate states $\ell=0$ and $\ell = n$. The vertical dotted line indicates marks the transition to the axial-FM phase. (b) $\Delta_{\Gamma}$ Lowest magnon energy at $\Gamma$, $E_\Gamma$, as a function of $B$ for several tube diameters $D$. (c) Dirac gap $\Delta_K$ as a function of $B$ for several $D$. $g \mu_B B_{s} \simeq 0.3$~meV for $D=4.1$~nm.}
    \label{fig:splitt}
\end{figure}

\textit{Magnon dispersion}---
Fig.~\ref{fig:2}(a) shows the magnon dispersion of the $(11,11)$ armchair nanotube, for the parameters given in the caption, in its three phases: radial ($B=0$), conical ($0 < B < B_s$), and axial-FM ($B\geq B_s$), with $B_s$ given in Eq.~\eqref{eq:bs}. In all three cases the magnon energies are real and positive, satisfying $E_n(k) = E_n(-k)$, so only $0 \le k \le \pi/a$ is shown. The dispersion relation exhibits a gap $\Delta_{\Gamma}$ at $\Gamma$  [see Fig.~\ref{fig:splitt}(b)] in the radial phase, which stems from the radial single-ion anisotropy. In the conical phase, the gap decreases with $B$ and shows a kink at $B \simeq B_s/2$ as a consequence of the crossing of the lowest levels. As the field reaches the saturation field $B_s$, the $\Gamma$ gap closes. In the axial-FM phase it reopens following a Zeeman shift linear with $B$. The resulting V-shaped field dependence near $B_s$ provides a direct experimental fingerprint of the conical-to-axial transition of the system, which could be proved by ferromagnetic resonance~\cite{shen2021multi}.

Fig.~\ref{fig:2}(b) shows the same spectra in the helical representation, where each $4\times4$ block of the magnon Hamiltonian is labeled by an angular index $\mu$. The highlighted bands belong to the two sectors involved in the Dirac crossing, $\mu_D=0$ (blue) and $\mu_{D'}=n-1$ (orange). In the radial phase the two $\mu_D$ bands cross at 
$k_D \simeq \frac{2\pi}{3a}$,
whereas the $\mu_{D'}$ bands remain gapped. In the axial-FM phase the roles are reversed: the crossing is formed by the $\mu_{D'}$ bands, and the $\mu_D$ bands are gapped. In the conical phase neither sector crosses, which opens a gap $\Delta_K$ at the Dirac crossings, [see Fig.~\ref{fig:splitt}(c)].

In the radial and axial-FM phases, two bands cross at $E \simeq 10$~meV, near $k=\frac{2\pi}{3a}$, i.e., the projection of the Dirac points of the honeycomb lattice onto the axis of an armchair tube. Curvature shifts this crossing slightly away from $2\pi/(3a)$. In an armchair tube one of the three nearest-neighbor bonds lies along the circumference and spans an azimuthal separation $|\varphi|=\frac{2\pi}{3n}$, whereas the other two have only half of their length along it with azimuthal separation $|\varphi|=\frac{\pi}{3n}$. This is the magnon analogue of the curvature-induced displacement of the Dirac points in carbon nanotubes~\cite{Kane1997,Laird2015}.

Key properties of the magnon dispersion, namely, the multiple band crossings, the gapping of the Dirac crossings, and the linear splitting of the doublets (See fig. \ref{fig:splitt}(a)), stem from the complex part of the exchange hopping of the magnon Hamiltonian. Since each spin is quantized along its classical direction $\bm{m}_i$ in a local frame, the hopping between neighboring sites $i$ and $j$ is proportional to $\bm{u}_i \cdot \bm{u}_j^* = (1 + \bm{m}_i \cdot \bm{m}_j) e^{i\chi_{ij}}$,
where $\bm{u}_i$ and $\bm{u}^*_i$ form the circular basis of transverse plane to $\bm{m}_i$ [see Eq.~\eqref{eq:scalar-prods}], where $\chi_{ij} = 2\arctan[\sin\theta \tan\varphi_{ij}/2] \simeq  \sin\theta \varphi_{ij}$ acts as a  Peierls phase for the HP boson hopping associated to exchange. (End Matter B). This phase vanishes in the radial phase, reduces to the pure gauge $e^{i\varphi_{ij}}$ in the axial-FM phase, and becomes non-trivial in the conical phase. There it acts as an effective vector potential in the HP bosons~\cite{schutz03, dugaev05}. Because the phase is odd in $\varphi_{ij}$, the phases cancel around every hexagon, so magnons feel no local magnetic field. However, denoting by $\chi_1$ and $\chi_2$ the phases acquired by a magnon state hopping between neighboring sites with azimuthal separation $\varphi_{1,2}=\frac{2\pi}{3n}, \frac{\pi}{3n}$, a magnon encircling the tube picks up a phase,
\begin{equation}
    \Phi = 2n (\chi_1 + \chi_2) \simeq 2\pi \sin\theta = 2\pi B/B_s, \quad 0 \leq B \leq B_s,
\end{equation}
which equals, modulo $2\pi$, the Berry phase of the local spin frame as it sweeps a cone of solid angle $2\pi(1-\sin\theta)$ around the tube. The tube is thus threaded by an effective Aharonov-Bohm flux~\cite{aharonov1959significance}, which the field tunes from zero in the radial phase to one flux quantum $\Phi = 2\pi$ in the axial-FM phase.

To make the effect of the magnetic flux on the magnon dispersion transparent, we make use of the helical representation and its correspondence with the axial one. In the helical representation, the magnon states are labeled in terms of the angular momentum number $\mu$, and the helical wavevector $q$. We now use the relation between the translational operator $\mathcal{T}$, with eigenvalues $e^{ik}$, and the screw ($U_{\mathrm{screw}}$) and rotation ($C_n$) operators, with eigenvalues $e^{iq}$ and $e^{i \frac{2\pi\mu}{n}}$, $\mathcal{T} = C_n^{-1} U_{\mathrm{screw}}^2$, to relate the helical and axial representations,
\begin{equation}
    q = \frac{\pi \mu}{n} + \frac{k}{2} + j\pi,\quad j = 0, 1.
\end{equation}
So, for each $\mu$ the two helical sectors $q$ and $q+\pi$ fold onto the same $k$. 
Whereas $\mu$ labels the  bands in both branches in  the helical representation [Fig.~\ref{fig:2}(b)],  we introduce $\ell = \mu + jn \mod{2n}$ to label the bands in the axial representation.
The axial representation exploiting $C_n$ symmetry therefore
contains twice as many bands as the helical representation:
the additional index $j=0,1$ reflects its four-site unit cell,
compared with the two-site unit cell of the helical representation.

The flux enters only through the combination $\ell + \Phi/2\pi$ (End Mater-D). This is exactly how an axial magnetic flux shifts the quantized angular momentum of electrons in carbon nanotubes~\cite{ajiki1993}. Consequently, after the field drives the system from the radial to the axial-FM phases, a full flux quantum is absorbed by relabeling $\ell \to \ell+1$. In the axial-FM, this reflects the winding of the local frames once around the tube.

\textit{Large pseudo-orbital magnetic moment of magnons}---Following \refcite{neumann20}, we define the magnetic moment of a magnon state along the tube axis as $m_z(\ell,k)\equiv -\frac{d E_\ell(k)}{d B}$ and split it into two contributions,
\begin{equation}
\label{eq:moment-decomp}
    m_z(\ell,k) = g\mu_B\,\mathcal{S}_z(\ell,k) + \mu_z(\ell,k).
\end{equation}
The first term arises from the explicit field-dependence of the magnon energies, which enters only through the Zeeman term, so the Hellmann-Feynman theorem gives $\partial E_\ell / \partial B|_{\theta} = - g \mu_B \mathcal{S}_z(\ell,k)$, with $\mathcal{S}_z\simeq-\sin\theta$ the change in the total spin along the axis carried by the magnon. The second term is  the orbital magnetic moment of magnons proposed in Ref.~\cite{neumann20}. We refer to it as the pseudo-orbital magnetic moment, since magnons carry no charge, so their orbital motion does not by itself produce a magnetic moment \cite{tang2026}.  

The $\mu_z$ contribution arises from the texture-dependence of the magnon energies implicitly driven by the field, which is given by
\begin{equation}
\label{eq:mu-texture}
    \mu_z(\ell,k) = -\left. \frac{\partial E_\ell(k)}{\partial\theta} \right|_{B} \,\frac{d\theta}{dB}.
\end{equation}
Two consequences follow: first, in the axial-FM phase, $\theta=\pi/2$ does not depend on $B$, so $\mu_z=0$ and magnons carry only their spin moment. Second, the conical texture is soft: $\theta$ rotates from the radial to the axial-FM phase within $g \mu_B B_s\simeq0.3$~meV, not far from $2KS$ [Eq.~\eqref{eq:bs}], whereas the magnon energies disperse with $\theta$ on the $JS$ energy scale. Hence $d \theta / dB \sim 1/B_s$ is large, and $\mu_z$ can greatly exceed $g\mu_B$.

We now discuss the pseudo-orbital moment of magnons at the $\Gamma$ point for $B\simeq 0$.
For the $(n,n)$ tube there are $2n$ such states, that we can label with $\ell$ that takes values from 0 to $2n-1$, modulo $2n$.  For $B=0$ the HP Hamiltonian is real, so that complex conjugation of an eigenstate with label $\ell$  gives a degenerate partner $\overline{\ell}=2n-\ell$. It is apparent that $\ell=0$ and $\ell=n$ are self-conjugate, and therefore, label non-degenerate states.

Expanding the quadratic HP Hamiltonian about the radial state,
the only terms linear in the axial field arise from the Peierls phases of the
number-conserving exchange hopping:
\begin{equation}
\mathcal{V}_{\rm linear}
=
-iJS\frac{B}{B_s}
\sum_{\langle ij\rangle}
\sin\varphi_{ij}
\left(a_i^\dagger a_j-a_j^\dagger a_i\right),
\label{eq:linear-field}
\end{equation}
where each nearest-neighbor bond is counted once.
Neglecting pairing and retaining the leading order in curvature,
its expectation value yields the zero-field pseudo-orbital
magnetic moment at $\Gamma$:
\begin{equation}
\mu_z(\ell,0)
\simeq
\sigma\frac{2\pi JS}{nB_s}
\frac{\sin(\pi\ell/n)}
{\sqrt{5+4\cos(\pi\ell/n)}},
\label{eq:orbital-moment-Gamma}
\end{equation}
Here $\sigma=-1$ ($+1$) denotes the acoustic (optical)
branch. 
$\mu_z$ is the response of the magnons circulating around the tube to the flux $\Phi$, analogous to the orbital moment of a charged particle in a Aharonov-Bohm ring. It is apparent that  the time reversal partners $\ell$ and $\overline{\ell}$ have opposite pseudo-orbital magnetic moments,  $\mu_z(\ell,0)=-\mu_z(\overline{\ell},0)$, 
accounting for splitting at finite $B$, in the conical phase, shown  in Fig.~\ref{fig:splitt}(a).  For $\ell \ll n$,  the pseudo-orbital magnetic moment scales linearly with  $\ell$ and is inversely proportional to the square of the tube radius $n^2$ for wide tubes, $\frac{\mu_z(\ell,0)}{g\mu_B}
\simeq \frac{2\pi^2 JS}{3B_s}\,\frac{\ell}{n^2}$,  indicating they are induced by   exchange $J$ and curvature $1/n$, in competition with radial anisotropy $K$. We also note that tubes with small $g \mu_B B_s$,  close to the critical line where the radial phase is not stable, will feature larger pseudo-orbital moments, but will only be stable for smaller values of $B$. The linear contribution of $B$ to the $\Gamma$  magnons of the radial phase energies vanishes for the non-doublet states, with $\ell=0,n$.

From Eq.~\eqref{eq:orbital-moment-Gamma}, we estimate the maximum value of the pseudo-orbital magnetic moments as
\begin{equation}
\label{eq:mu-max}
|\mu_z|_{\max} \simeq \frac{\pi JS}{n\,B_s},
\end{equation}
which, for the  CrI$_3$ parameters used here, amounts to $\simeq 6.2 \mu_B$, about three times the largest spin contribution, in agreement with Eq.~\eqref{eq:mu-max}.

At $B\simeq B_s/2$, i.e., $\Phi\simeq\pi$, the levels regroup into nearly degenerate pairs $(\ell,-\ell-1)$, which produces the crossings in the middle of Fig.~\ref{fig:splitt}(a). This spectral flow is analogous to the rearrangement of angular-momentum levels of a charged particle in an Aharonov-Bohm ring as the threading flux is varied. At $B=B_s$, the full flux quantum is absorbed by the relabeling $\ell\to\ell+1$, so the doublets form again between $\ell$ and $-\ell-2$, and the slopes change abruptly. In the axial-FM phase the flux is frozen, $\mu_z=0$, and all levels shift with nearly the same slope, $|m_z|\simeq g\mu_B$.

\paragraph{Discussion and Conclusions---}
We have shown that magnons in $\mathrm{CrI}_3$ nanotubes provide a fingerprint of their radial and conical magnetic order. The underlying mechanism is geometric: the conical texture, arising from the interplay between curvature, radial single-ion anisotropy, and an axial field, acts on magnons as an effective Aharonov-Bohm flux threading the tube, which the field tunes continuously from zero in the radial phase to one flux quantum in the axial-FM phase. This flux gaps the Dirac crossings in the conical phase and endows magnons with a pseudo-orbital magnetic moment that splits modes of opposite circulation, reaching $\simeq 6 \mu_B$ for an $(11,11)$ armchair $\mathrm{CrI}_3$ tube. 
This value is 20 times larger than predictions for planar systems  \cite{neumann20,jeon2026} and are relatively  closer to  electronic orbital moments 
measured in carbon nanotubes~\cite{Minot04,Island2018}.
Although this moment stems from the field-induced texture reorientation, is associated with the winding of magnons around the tube. By contrast, in the axial-FM phase magnons carry only their spin moment. Together with the softening of the lowest mode at $B_s \simeq 2.6$~T, these features could be resolved by electrically detected spin resonance in the shell nanotube\cite{pei2012} and inelastic electron tunneling spectroscopy~\cite{klein18,Ghazaryan2018} on individual tubes, taking advantage of the fact that CrI$_3$ tubes are encapsulated in multiwall carbon nanotubes\cite{caha2025}. In addition, since the flux is fixed by the solid angle swept by the magnetization, we expect the same phenomenology in chiral tubes and in other curved magnets with radial textures.

\vspace{1cm}

\paragraph{Acknowledgments.---}
D.S.R.
acknowledges funding from grant PID2022-141712NB-C22, funded by MICIU/AEI/10.13039/501100011033 and FEDER, UE.
J.F.-R. 
acknowledges financial support from 
Swiss National Science Foundation Sinergia grant PiMag,
from Generalitat Valenciana (Grants No.~Prometeo2021/017 and  No.~MFA/2022/045),
and from MICIN-Spain (Grants No.~PID2022-141712NB-C22 and No.~PRTR-C17.I1).
We acknowledge the use of ChatGPT (OpenAI) and Claude (Anthropic) as a research aid for theoretical discussions and manuscript editing. We have carefully revised all the results. Accordingly, we take full responsibility for the present material.

\bibliography{refs}

\onecolumngrid
\vspace{0.5em}
\begin{center}
  \textbf{\large End Matter}
\end{center}
\vspace{0.3em}
\twocolumngrid

\setcounter{equation}{0}
\renewcommand{\theequation}{EM\arabic{equation}}

\newcounter{appsec}
\renewcommand{\theappsec}{\Alph{appsec}}
\newcommand{\AppSec}[2]{%
  \refstepcounter{appsec}%
  \section*{\theappsec. #2}%
  \label{#1}%
}

\AppSec{ap:A}{Magnon Hamiltonian}
Here we sketch the main steps of the derivation of Eq.~\eqref{eq:magnon-ham-final}. We first express each spin in a local frame whose third axis is aligned with the classical direction $\bm{m}_i$ [Eq.~\eqref{eq:param-magn}]. For the canted radial configuration, we choose
\begin{equation}
\label{eq:local-triad}
    \bm{e}_{1,i} = -\bm{e}_{\varphi,i} \times \bm{m}_i, \quad
    \bm{e}_{2,i} = -\bm{e}_{\varphi,i}, \quad
    \bm{e}_{3,i} = \bm{m}_i,
\end{equation}
where $\bm{e}_{r,i} = (0, \cos\varphi_i, \sin\varphi_i)$, $\bm{e}_{\varphi,i} = (0, -\sin\varphi_i, \cos\varphi_i)$, and $\bm{e}_z = (1,0,0)$ are the cylindrical unit vectors at site $i$. Here $\varphi_i$ is the azimuthal angle of site $i$ around the tube axis, and $\theta$ is the canting angle, measured from the transverse plane toward the tube axis.

Defining $\bm{u}_i = \bm{e}_{1,i} + i \bm{e}_{2,i}$, the spin operator reads
\begin{equation}
\label{eq:local-frame-spin}
    \bm{S}_i = \frac12 \left( \tilde{S}^-_i \bm{u}_i + \tilde{S}^+_i \bm{u}^*_i \right) + \tilde{S}^z_i \bm{m}_i,
\end{equation}
where $\tilde{S}^{\pm}_i = \tilde{S}^x_i \pm i \tilde{S}^y_i$ are the ladder operators in the local frame. We then apply the HP transformation, $\tilde{S}^z_i = S - a^\dagger_i a_i$, $\tilde{S}^+_i \simeq \sqrt{2S}\, a_i$, and $\tilde{S}^-_i \simeq \sqrt{2S}\, a^\dagger_i$, retaining terms up to quadratic order in the bosonic operators. The linear terms vanish when the canting angle satisfies
\begin{equation}
\label{eq:Bs}
    \sin\theta = \frac{B}{B_s}, \quad g \mu_B B_s = S(2K - J\Gamma_n),
\end{equation}
where $B_s$ is the saturation field and $\Gamma_n = 3 - \cos\frac{2\pi}{3n} - 2\cos\frac{\pi}{3n}$ is the curvature-induced exchange cost for an $(n,n)$ tube. For $B \geq B_s$ the spins are fully aligned with the tube axis, $\theta = \pi/2$.

The resulting magnon Hamiltonian depends on the tube curvature only through the scalar products
\begin{equation}
\label{eq:scalar-prods}
\begin{split}
    \bm{m}_i \cdot \bm{m}_j &= \sin^2\theta + \cos^2\theta \cos\varphi_{ij}, \\
    \bm{u}_i \cdot \bm{u}^*_j &= \cos^2\theta + (1 + \sin^2\theta) \cos\varphi_{ij} + 2i \sin\theta \sin\varphi_{ij}, \\
    \bm{u}_i \cdot \bm{u}_j &= \cos^2\theta\, (1 - \cos\varphi_{ij}),
\end{split}
\end{equation}
with $\varphi_{ij} = \varphi_j - \varphi_i$, together with $\bm{m}_i\cdot\bm{e}_{r,i} = \cos\theta$ and $\bm{u}_i\cdot\bm{e}_{r,i} = -\sin\theta$ for the SIA term.

Two effects of the curvature follow directly from Eq.~\eqref{eq:scalar-prods}. First, the product $\bm{u}_i \cdot \bm{u}^*_j$, which enters through the exchange term of Eq.~\eqref{eq:spin-ham}, is real in the radial phase and reduces to a pure gauge phase, $2e^{i\varphi_{ij}}$, in the axial-FM phase. By contrast, in the conical phase its imaginary part yields a complex hopping between nearest neighbors that acts as an effective curvature-induced DMI. Second, anomalous terms $a^\dagger_i a^\dagger_j$ arise both from the radial SIA, onsite and with a strength proportional to $\sin^2\theta$, and from the curvature-induced noncollinearity in the exchange term, between nearest neighbors. Hence, the magnon number is not conserved and the classical ground state is not an eigenstate of the Hamiltonian.

Exploiting the translational invariance along the tube axis, we Fourier transform the bosonic operators using the translational unit cell of an $(n,n)$ nanotube, which contains $N_c = 4n$ sites,
\begin{equation}
    a_{i,R} = \frac{1}{\sqrt{N_{\mathrm{uc}}}} \sum_{k} e^{i k R}\, a_{i,k}.
\end{equation}
Here $a_{i,R}$ is the magnon annihilation operator at site $i$ of unit cell $R$, $N_{\mathrm{uc}}$ is the number of unit cells, and $k$ is the wave vector along the tube axis. In terms of the Nambu spinor $\psi_k = ( a_{1,k}, \dots, a_{N_c,k}, a^\dagger_{1,-k}, \dots, a^\dagger_{N_c,-k} )^{T}$, the magnon Hamiltonian reads
\begin{equation}
\label{eq:magnon-ham-quadratic}
    H = E_{\mathrm{cl}} + E_{\mathrm{corr}} + \frac12 \sum_{k} \psi^\dagger_{k}\, h(k)\, \psi_k,
\end{equation}
where $E_{\mathrm{cl}}$ is the classical ground-state energy, $E_{\mathrm{corr}}$ collects the $c$-number terms generated by reordering the magnon operators into Nambu form, and $h(k)$ is the Bogoliubov-de Gennes (BdG) matrix
\begin{equation}
\label{eq:bdg-matrix}
    h(k) =
    \begin{pmatrix}
        \mathcal{A}(k) & \mathcal{B}(k) \\
        \mathcal{B}^\dagger(k) & \mathcal{A}^*(-k)
    \end{pmatrix},
\end{equation}
with blocks
\begin{equation}
\label{eq:bdg-blocks}
\begin{split}
    \mathcal{A}_{ij}(k) ={}& \delta_{ij} \Bigl[ JS \!\!\sum_{j' \in \mathrm{NN}(i)}\!\! \bm{m}_i \cdot \bm{m}_{j'} \\
    & + KS \left( 2\cos^2\theta - \sin^2\theta \right) + g \mu_B B \sin\theta \Bigr] \\
    & - \frac{S}{2} J_{ij}(k)\, \bm{u}_i \cdot \bm{u}^*_j, \\
    \mathcal{B}_{ij}(k) ={}& -\frac{S}{2} J_{ij}(k)\, \bm{u}_i \cdot \bm{u}_j - \delta_{ij} KS \sin^2\theta.
\end{split}
\end{equation}
Here $i$ and $j$ label sites within the unit cell, and $J_{ij}(k)= J\sum_{j'}e^{2\pi i k (R_{j'}-R_i)}$ is the Fourier transform of the exchange coupling, where the sum runs over all nearest neighbors $j'$ of site $i$ that are equivalent to $j$ up to a lattice translation, and $R_{j'}-R_i$ is the separation between their unit cells.

The magnon energies and eigenvectors are obtained through a paraunitary transformation\cite{colpa1978} following Colpa's method \cite{colpa1978, toth2015}, $\psi_k = P(k)\, \chi_k$, which maps the magnon operators onto Bogoliubov quasiparticles while preserving the bosonic commutation relations. Writing $\psi_k = (a_k, a^\dagger_{-k})^{T}$ and $\chi_k = (\alpha_k, \alpha^\dagger_{-k})^{T}$, with $\alpha_k = (\alpha_{1,k}, \dots, \alpha_{N_c,k})^{T}$, this reads explicitly
\begin{equation}
    \begin{pmatrix}
        a_{k} \\ a^\dagger_{-k}
    \end{pmatrix} =
    \begin{pmatrix}
        U(k) & V^*(-k) \\
        V(k) & U^*(-k)
    \end{pmatrix}
    \begin{pmatrix}
        \alpha_{k} \\ \alpha^\dagger_{-k}
    \end{pmatrix},
\end{equation}

\AppSec{ap:B}{Berry phase of magnons around the tube}
The exchange term of the magnon Hamiltonian, Eq.~\eqref{eq:magnon-ham-quadratic}, contains a complex hopping [see Eq.~\eqref{eq:bdg-blocks}], whose phase can be interpreted as a Peierls phase. In the noncollinear texture, each spin is quantized along the classical direction $\bm{m}_i$, so the local frames of neighboring spins are rotated with respect to each other, and a magnon hopping between them acquires a geometric phase. Since the hopping of the exchange terms are proportional to $\bm{u}_i \cdot \bm{u}_j$, we can put Eq.~\eqref{eq:scalar-prods} in terms of a real amplitude accompanied by this geometric phase,
\begin{equation}
\label{eq:peierls}
\begin{split}
    \tfrac12\, \bm{u}_i\cdot\bm{u}_j^* &= \tfrac12 \left(1+\bm{m}_i\cdot\bm{m}_j\right) e^{i\chi_{ij}}, \\
    \chi_{ij} &= 2\arctan\!\left[\sin\theta\, \tan\frac{\varphi_{ij}}{2}\right] \simeq \varphi_{ij}\sin\theta = \varphi_{ij} \frac{B}{B_s} ,
\end{split}
\end{equation}
where the last expression holds for $|\varphi_{ij}| \ll 1$ for neighboring sites $i$, and $j$ and $0\leq B \leq B_s$. In an $(n,n)$ armchair tube, the nearest-neighbor azimuthal separations are $|\phi_{ij}| = \frac{2\pi}{3n}, \frac{\pi}{3n}$. For the tubes considered here ($n \geq 11$), $|\varphi_{ij}| \leq 0.19$, and the leading correction changes $\chi_{ij}$ by less than $0.3\%$.

Consequently, the Peierls phase $\chi_{ij}$ is the lattice form of the Berry connection of magnons in a textured ferromagnet~\cite{dugaev05} and acts as a curvature-induced DMI. It vanishes in the radial state and is a pure gauge in the axial-FM state, which originates from the rotations of the local frames around the tube axis. Since $\chi_{ij}$ is an odd function of $\varphi_{ij}$ alone, the phases on opposite bonds cancel around every hexagon, whereas a magnon looping once around the circumference acquires
\begin{equation}
\label{eq:flux}
    \Phi = \sum_{(ij)\in\mathcal{C}} \chi_{ij}  = 2n(\chi_1 + \chi_2) \simeq 2\pi - \Omega = 2\pi\sin\theta  ,
\end{equation}
where $\mathcal{C}$ is a closed path winding once around the tube, $\chi_1$ and $\chi_2$ are the acquired phases across the azimuthal separations, $\varphi_1 = \frac{2\pi}{3n}$ and $\varphi_2=\frac{\pi}{3n}$, and $\Omega = 2\pi(1-\sin\theta)$ is the solid angle of the cone swept by $\bm{m}_i$. The tube is thus threaded by an effective Aharonov--Bohm flux $\Phi \simeq 2\pi B/B_s$ tuned by the field, and the modes $a_i \propto e^{i\mu\varphi_i}$, with integer $\mu$, depend on $\mu$ through $\mu + \Phi/2\pi$, where the flux is shared equally among the $n$ $\mu$-sectors.

\AppSec{ap:C}{Helical Hamiltonian}
Here we outline the construction of the LSWT magnon Hamiltonian in a basis adapted to the rotational and helical symmetries of the tube, following Refs.~\cite{izumida2016, izumida2017topology, okuyama2019topological}. Instead of the translational unit cell, we use for an $(n,n)$ armchair nanotube the two-site cell spanned by $\bm{C}_h/n$ and the helical vector $\bm{R}=\bm{a}_1$, where $\bm{C}_h=n(\bm{a}_1+\bm{a}_2)$ is the chiral vector and $\bm{a}_1$, $\bm{a}_2$ are the primitive vectors of the honeycomb lattice that is rolled up to form the tube. In this basis the Hamiltonian is block diagonal in the angular quantum number $\mu$ and the helical wave vector $q$, and the magnon dispersion follows from the Bogoliubov diagonalization of a $4\times4$ BdG matrix for each pair $(\mu,q)$; we refer to this method as the helical representation.

We write the spin positions as $\bm{r}=\nu\,\bm{C}_h/n+l\,\bm{R}+\bm{s}_\tau$. Here $\nu=0,\dots,n-1$ labels the position around the circumference, the integer $l$ counts steps of $\bm{R}$ along the helix, each of which advances $a/2$ along the tube axis and rotates by $\pi/n$ about it ($a$ is the lattice constant of the honeycomb lattice), and $\bm{s}_\tau$ is the position of sublattice $\tau=A,B$ within the cell. With the spins expressed in local frames that rotate with the lattice sites, the couplings depend only on the relative position of the sites, and the Hamiltonian is block diagonalized by the Fourier transform
\begin{equation}
    a_{\nu l \tau}=\frac{1}{\sqrt{N_{\mathrm{uc}}}}\sum_{\mu,q}e^{i(2\pi\mu\nu/n+ql)}\,a_{\bm{k}\tau},
\end{equation}
where $\bm{k}=(\mu,q)$ with $\mu=0,\dots,n-1$ and $q\in[-\pi,\pi)$, and $N_{\mathrm{uc}}$ is the number of cells. Up to a constant, this yields $\mathcal{H}=\frac{1}{2}\sum_{\bm{k}}\Psi^\dagger_{\bm{k}}\,h(\bm{k})\,\Psi_{\bm{k}}$, with $\Psi_{\bm{k}}=(a_{\bm{k}A},a_{\bm{k}B},a^\dagger_{-\bm{k}A},a^\dagger_{-\bm{k}B})^{T}$, $-\bm{k}\equiv(-\mu \bmod n,-q)$, and
\begin{equation}
    h(\bm{k}) =
    \begin{pmatrix}
        \mathcal{A}(\bm{k}) & \mathcal{B}(\bm{k}) \\
        \mathcal{B}^\dagger(\bm{k}) & \mathcal{A}^*(-\bm{k})
    \end{pmatrix}.
\end{equation}
In sublattice space, the blocks are
\begin{equation}
    \mathcal{A}(\bm{k}) =
    \begin{pmatrix}
        \varepsilon_0 & f_{\bm{k}} \\
        f^*_{\bm{k}} & \varepsilon_0
    \end{pmatrix},
    \quad
    \mathcal{B}(\bm{k}) =
    \begin{pmatrix}
        \kappa & g_{\bm{k}} \\
        g^*_{\bm{k}} & \kappa
    \end{pmatrix}.
\end{equation}
The onsite energy of subblock $\mathcal{A}$ is
\begin{equation}
\label{eq:eps0}
\begin{split}
    \varepsilon_0 ={}& JS\left[\Gamma_n\sin^2\theta+\cos\varphi_n+2\cos\frac{\varphi_n}{2}\right] \\
    & +KS\left(2-3\sin^2\theta\right)+ g \mu_B B\sin\theta,
\end{split}
\end{equation}
where $\varphi_n=2\pi/3n$. The onsite energy of the pairing terms is given by
\begin{equation}
\label{eq:kappa}
    \kappa = -KS \sin^2\theta.
\end{equation}

In the normal terms of $\mathcal{A}$, $A$ site in cell $(\nu,l)$ couples to the $B$ sites in cells $(\nu,l)$, $(\nu,l-1)$, and $(\nu-1,l+1)$, at azimuthal separations $\varphi_n$, $-\varphi_n/2$, and $-\varphi_n/2$, respectively. Fourier transforming and using Eq.~\eqref{eq:peierls} and Eq.~\eqref{eq:flux}, the normal hopping reads off
\begin{equation}
\label{eq:fk-2}
    f_{\bm{k}} = -JS e^{i\chi_1} \left[ 
    t_1 + 2t_2e^{-i\pi(\mu + \Phi/2\pi)/n} \cos\left(q - \frac{\pi \mu}{n}\right)
    \right],
\end{equation}
with $\bm{k} = (\mu,q)$ and $\chi_1 = \chi(\varphi_n)$ the Peierls phase between sites with azimuthal separation $\varphi_n$. We define the absolute value of the hoppings between the two kinds of nearest-neighbor bonds as $t_1 = t(\varphi_n)$ and $t_2 = t(\frac{\varphi_n}{2})$, with 
\begin{equation}
\label{eq:tfuncs}
    t(\varphi) = \frac12 \left( 1 + \bm{m}_i \cdot \bm{m}_j \right) = 1 - \cos^2\theta \sin^2\frac{\varphi}{2}.
\end{equation}
The anomalous term $g_{\bm{k}}$ is given by
\begin{equation}
\label{eq:gk}
g_{\bm{k}} = -\frac{JS}{2} \left[ 
    g(\varphi_n) + 2g(-\frac{\varphi_n}{2})e^{-i\pi\mu/n} \cos\left(q - \frac{\pi \mu}{n}\right)
\right],
\end{equation}
with 
$g(\varphi)=\cos^2\theta\,(1-\cos\varphi)$.

\AppSec{ap:D}{Correspondence between helical and axial representations}
The magnon dispersion exhibits two remarkable features. The Dirac crossing is gapless in the $\mu=0$ sector in the radial phase and in the $\mu=-1$ sector in the axial-FM phase, and is gapped across the conical phase. In addition, the doublets that are degenerate in the radial phase split in linearly and in opposite directions with the field, reminiscent of the Zeeman splitting of a magnetic moment. Both effects become transparent once the helical Hamiltonian is expressed in terms of the axial wave vector making use of the helical-axial correspondence.

The spin positions in the helical representations are $\bm{r} = \nu \bm{C}_h/d + \bm{R} l + \bm{s}_\tau$. So, the azimuth coordinate of site $(\nu, l)$ is $\varphi_{\nu,l} = \frac{2\pi}{n}\nu + \frac{\pi}{n}l$, and the axial coordinate $z_l = \frac{la}{2}$. The screw operation $U_{\mathrm{screw}}$ applies a  rotation around the tube axis by $\frac{\pi}{n}$ and a translation by $\frac{a}{2}$ along the axis, which yields
\begin{equation}
\label{eq:q-to-kmu}
    q = \frac{\pi}{n} \mu + \frac{k}{2} + j\pi, \quad j=0,1,
\end{equation}
so each pair of helical wave vectors $q$ and $q+\pi$ with the same angular momentum $\mu$ fold onto the same axial wave vector $k$.

The axial wave vectors are decoupled from the effect of the flux, so, making use of this correspondence in Eq~\eqref{eq:fk-2}, we combine the angular momentum $\mu$ and the sector index $j$ into angular index $\ell$,
\begin{equation}
\label{eq:ang-axial-index}
    \ell = \mu + jn \pmod{2n}, \quad j=0,1,
\end{equation}
and Eq.~\eqref{eq:fk-2} can be put into the form
\begin{equation}
\label{eq:fk-3}
    f(\mu,k) = -JS e^{i\chi_1} 
    \left[
    t_1 + 2t_2 e^{-i\pi(\ell + \Phi/2\pi)/n} \cos\left(\frac{k}{2}\right)
    \right],
\end{equation}
where $\cos\left(\frac{k}{2}\right) \geq 0$ in the first Brillouin zone.

Neglecting the pairing terms, in the radial phase ($\Phi=0$) the exchange hopping of $\mu=0$ is real up to a gauge phase, which guarantees that there exists $k$ where $f_{\bm{k}} = 0$, i.e, a gapless band. In the axial-FM phase ($\Phi = 2\pi$), $\mu = -1$ satisfies the same condition, which explains the switch of the gapless band from $\mu=0$ to $\mu=-1$. In addition, in the radial and axial-FM phases the pairing terms do not contribute to opening the gap. By contrast, in the conical phase the exchange complex cannot be converted into a real value up to a gauge phase, so there are no nodes, and all states remain gapped.

\nolinenumbers

\end{document}